\documentclass[12pt,tightenlines,eqsecnum,floats,aps,amsmath,amssymb,nofootinbib,prd,showpacs]{revtex4}

\usepackage{setspace}
\usepackage{amsmath,amssymb,amsfonts,amsthm}
\usepackage{graphicx}
\usepackage{enumerate}

\begin{document}
\title{Quantum unitary evolution of linearly polarized $\mathbb{S}^1\times \mathbb{S}^2$ and $\mathbb{S}^3$ Gowdy
 models coupled to massless scalar fields}

\author{J. Fernando \surname{Barbero G.}}
\email[]{fbarbero@iem.cfmac.csic.es} \affiliation{Instituto de
Estructura de la Materia, CSIC, Serrano 123, 28006 Madrid, Spain}
\author{Daniel  \surname{G\'omez Vergel}}
\email[]{dgvergel@iem.cfmac.csic.es} \affiliation{Instituto de
Estructura de la Materia, CSIC, Serrano 123, 28006 Madrid, Spain}
\author{Eduardo J. \surname{S. Villase\~nor}}
\email[]{ejsanche@math.uc3m.es} \affiliation{Grupo de
Modelizaci\'on y Simulaci\'on Num\'erica, Universidad Carlos III
de Madrid, Avda. de la Universidad 30, 28911 Legan\'es, Spain}
\affiliation{Instituto de Estructura de la Materia, CSIC, Serrano
123, 28006 Madrid, Spain}

\date{November 12, 2007}

\begin{abstract}
The purpose of this paper is to study in detail the problem of
defining unitary evolution for linearly polarized
$\mathbb{S}^1\times\mathbb{S}^2$ and $\mathbb{S}^3$ Gowdy models (in
vacuum or coupled to massless scalar fields). We show that in the Fock
quantizations of these systems no choice of acceptable complex structure leads
to a unitary evolution for the original variables. Nonetheless,
unitarity can be recovered by suitable redefinitions of the basic
fields. These are dictated by the time dependent conformal factors
that appear in the description of the standard deparameterized form
of these models as field theories in certain curved backgrounds. We also
show the unitary equivalence of the Fock quantizations obtained from the
$SO(3)$-symmetric complex structures for which the dynamics is unitarily
implemented.
\end{abstract}


\maketitle

\section{Introduction}{\label{Intro}}

Gowdy models \cite{Gowdy:1971jh, Gowdy:1973mu} are interesting
$U(1)\times U(1)$ symmetry reductions of $(1+3)$-general relativity
that have been used for a number of years as test beds for quantum
gravitational techniques
\cite{Cortez,Misner,Berger:1973,Berger:1975kn,Ashtekar:1996bb,Mena:1997,Pierri:2000ri,Corichi:2002vy,
Torre:2002xt,BarberoG.:2006zw,Corichi:2006xi,Corichi:2006zv,Torre:2007zj,Mena:2007}.
They are receiving a lot of attention these days as the next arena
to test loop quantum gravity. In this sense they are the natural
continuation of the minisuperspace reductions considered so far in
loop quantum cosmology. The fact that they can be exactly solved
classically and admit a simple enough Hamiltonian description after
deparameterization (see \cite{BarberoG.:2007} for a rigorous
classical treatment of these models coupled to matter scalar fields)
makes them very attractive from this point of view.

One of the most striking features of the Fock space quantization for
the Gowdy $\mathbb{T}^3$ model is the impossibility of defining a
unitary quantum evolution operator when the system is written in the
natural field variables usually employed to describe it as a
$(1+2)$-dimensional field theory \cite{Corichi:2002vy,
Torre:2002xt}. Nevertheless this is not an unsurmountable problem
because it is possible to introduce a time-dependent field
redefinition that leads to a unique (up to unitary equivalence)
quantization, with unitary time evolution, when one demands
invariance under the residual $U(1)$ symmetry
\cite{Corichi:2006xi,Corichi:2006zv,Mena:2007}.

The purpose of this paper is to study the problem of the unitary
implementation of dynamics as a natural extension of the previous
literature devoted to the vacuum Gowdy $\mathbb{T}^{3}$ model. The
proposed generalization is two-fold. On one hand we will deal with
the remaining topologies admissible for the compact Gowdy models,
i.e. the three-handle $\mathbb{S}^1\times\mathbb{S}^2$ and the
three-sphere $\mathbb{S}^3$. On the other hand we will consider the
addition to the models of certain matter fields --massless scalars--
symmetric under spatial isometries \cite{BarberoG.:2005ge}.  Here we
will closely rely on the results of \cite{BarberoG.:2007}.

Our starting point is the interpretation of the compact Gowdy models
in the different topologies as scalar field theories in very
specific curved backgrounds. As shown in \cite{BarberoG.:2007} all
these models can be reinterpreted as given by the evolution of
massless scalar fields in some geometric backgrounds that are
conformally equivalent to the simplest metrics that can be defined
on each of the relevant $(1+2)$-dimensional space-time manifolds. In
particular, for the $\mathbb{T}^3$ case, the metric is just the flat
metric on $(0,\infty)\times\mathbb{T}^2$ whereas in the
$\mathbb{S}^1\times\mathbb{S}^2$ and $\mathbb{S}^3$ examples the
metric is the Einstein metric\footnote{In the $\mathbb{S}^3$ case
this is a non-trivial result \cite{BarberoG.:2007} that is found by
carefully considering the relevant regularity condition for the
fields.} on $(0,\pi)\times\mathbb{S}^2$. The corresponding conformal
factors are simple functions of $t$ ($t$ and $\sin t$ respectively).
We will use here this description to gain useful insights on the
problem of the unitary implementability of quantum time evolution.

As a first step towards this goal we show, by a direct argument,
that no choice of $SO(3)$-invariant complex structure leads to
unitary quantum evolution in terms of the variables in which these
systems are naturally written\footnote{In the vacuum Gowdy
$\mathbb{T}^3$ model this result is a direct corollary of the
uniqueness theorem appearing in \cite{Mena:2007}. Here we will
concentrate on the remaining topologies. Similar results for Gowdy
$\mathbb{T}^3$ coupled to massless scalar fields can be derived by
our methods in a straightforward way.}. The way out of this
seemingly unavoidable obstruction to quantization is to introduce a
time-dependent field redefinition as in \cite{Corichi:2006xi}; in
fact, by a simple re-scaling of the scalar fields involving
\textit{precisely} the conformal factors mentioned above we can get
a well defined and unitary quantum evolution not only in the
$\mathbb{T}^3$ model but for the other topologies as well. A way to
understand what is going on is to realize that the singular behavior
introduced by the conformal factors is translated, in terms of the
redefined fields, into the behavior of a singular, time-dependent,
potential term for the re-scaled fields. Time evolution can now be
implemented unitarily as a direct consequence of the fact that, in
spite of being singular at some instants of time, these potentials
are sufficiently well behaved as functions of the time variable in a definite
sense that will be explained below. We also show the uniqueness --modulo unitary
equivalence-- of the Fock quantizations that allow the unitary implementation
of the dynamics.

\bigskip

The paper is organized as follows. After this introduction we will
study in section \ref{PhaseSpace} the canonical and covariant phase
space descriptions of $\mathbb{S}^1\times\mathbb{S}^2$ and
$\mathbb{S}^3$ Gowdy models coupled to massless scalar fields, as
well as their classical dynamics. This is done by writing the field
equations with the help of a certain background metric. We also
discuss the appropriate mode decomposition of the fields. Section
\ref{FockQuantization} is devoted to several issues related to the
Fock quantizations of these systems. In particular, we obtain
different (in general unitarily nonequivalent) Fock representations
for the canonical commutation relations characterized by a
two-parameter family of $SO(3)$-invariant complex structures.
Section \ref{UnitarityEvolution} deals specifically with the
discussion of unitarity for the topologies considered in the paper.
Whereas it is not possible to implement in a unitary way the linear
symplectic transformation associated to the time evolution for the
original variables, we show that a suitable re-scaling of the fields
dictated by the conformal factor of the background metric leads to
unitarity. This quantization is unique up to unitary equivalence.
We also show here that despite having a well-defined and
unitary quantum dynamics the action of the Hamiltonian operator is
not defined on the Fock vacuum. This result is analogous to the one found
by the authors of \cite{Corichi:2006xi} in the $\mathbb{T}^3$ case.
We end the paper in section \ref{conclusions} with several comments and
a discussion of the results.

\section{Reduced phase space and classical dynamics}{\label{PhaseSpace}}

We discuss in this section the reduced phase space and the classical
evolution for the Gowdy models --in vacuum or coupled to massless
scalars-- corresponding to the $\mathbb{S}^1\times \mathbb{S}^2$ and
$\mathbb{S}^3$ topologies. The dynamics of the local degrees of
freedom that parameterize the reduced phase space in both cases
\cite{BarberoG.:2007} can be described by the same simple field
equations\footnote{In the following we will not consider the
dynamics of the global modes because it is irrelevant to the quantum
unitarity issues that we want to discuss in the paper. They can be
quantized  in  a straightforward way in terms of standard position
and momentum operators with dense domain in $L^{2}(\mathbb{R})$.}.
These can be written as wave equations with the help of a certain
auxiliary globally hyperbolic space-time background $((0,\pi)\times
\mathbb{S}^2,\mathring{g}_{ab} )$ and an extra symmetry condition:
invariance under the diffeomorphisms generated by a Killing vector
field $\sigma^a$ of $\mathring{g}_{ab}$. Explicitly the metric
$\mathring{g}_{ab} $ is
\begin{eqnarray}
\mathring{g}_{ab}=\sin^2t[-(\mathrm{d}t)_a(\mathrm{d}t)_b+\gamma_{ab}]\,,\label{g0}
\end{eqnarray}
where $\gamma_{ab}$ is the round unit metric on the 2-sphere
$\mathbb{S}^2$. Using spherical coordinates $(\theta,\sigma)\in
(0,\pi)\times (0,2\pi)$ on $\mathbb{S}^2$
\begin{eqnarray*}
\mathring{g}_{ab}&=&\sin^2t[-(\mathrm{d}t)_a(\mathrm{d}t)_b
+(\mathrm{d}\theta)_a(\mathrm{d}\theta)_b+\sin^2\theta(\mathrm{d}\sigma)_a(\mathrm{d}\sigma)_b]\,,
\end{eqnarray*}
and the Killing field $\sigma^a$ is simply $(\partial/\partial\sigma)^a$.

The field equations can be derived, by imposing the additional
symmetry condition $\mathcal{L}_\sigma\phi_i=0$ ($i=0,\ldots,N$) on
the solutions, from the action
\begin{eqnarray}
S(\phi_i)&=&-\frac{1}{2}\sum_{i=0}^N\int_{[t_0,t_1]\times
\mathbb{S}^2} |\mathring{g}|^{1/2}
\mathring{g}^{ab}(\mathrm{d}\phi_i)_a(\mathrm{d}\phi_i)_b\label{accion}
\\
&=&\frac{1}{2}\sum_{i=0}^N\int_{t_0}^{t_1}
\mathrm{d}t\int_{\mathbb{S}^2}|\gamma|^{1/2}\sin
t\,\Big(\dot{\phi}_i^2+\phi_i\,\Delta_{\mathbb{S}^2}\phi_i\Big)\,.\nonumber
\end{eqnarray}
Here and in the following $\dot{\phi}:=\partial\phi/\partial t$,
$\Delta_{\mathbb{S}^2}$ is the Laplace-Beltrami operator on the
round sphere $\mathbb{S}^2$, and $\mathcal{L}$ denotes the Lie
derivative. As shown in \cite{BarberoG.:2007} one of the scalar
fields, say $\phi_0$, encodes the local gravitational degrees of
freedom and the remaining ones, $\phi_i,\,i=1,\ldots,N$, describe
the matter modes added to the Gowdy models. As we can see they
completely decouple in this description\footnote{Notice that, in
spite of the apparent simplicity of the reduced phase space
description, the full $(3+1)$-dimensional metric that solves the
Einstein-Klein-Gordon equations depends both on the gravitational
and scalar modes in a non-trivial way \cite{BarberoG.:2007}.}
because, at variance with the $\mathbb{T}^3$ case, no extra
constraint remains. Owing to this fact we omit  in the following the
$i$ index whenever it is not necessary to explicitly separate
gravitational and matter modes.

\subsection{Canonical and covariant phase spaces}

We start by explicitly writing the linear space of smooth and
symmetric real solutions to the massless Klein-Gordon equation of
motion as
\begin{eqnarray}
\mathcal{S}&:=&\{\phi\in
C^\infty((0,\pi)\times\mathbb{S}^2;\mathbb{R})\,|\,\mathring{g}^{ab}\mathring\nabla_a\mathring\nabla_b\phi=0;\,
\mathcal{L}_\sigma\phi=0\}\label{solspace}\\
&=&\{\phi\in
C^\infty((0,\pi)\times\mathbb{S}^2;\mathbb{R})\,|\,\ddot{\phi}+\cot
t \dot{\phi}-\Delta_{\mathbb{S}^2}\phi=0;\,
\mathcal{L}_\sigma\phi=0\}\nonumber
\end{eqnarray}
endowed with the (weakly) symplectic structure $\Omega$ induced by
(\ref{accion})
\begin{equation}
\Omega(\phi_1,\phi_2):=\sin t\int_{\mathbb{S}^2}
|\gamma|^{1/2}\imath^*_t\Big(\phi_2\dot{\phi}_1-\phi_1\dot{\phi}_2\Big)\,.
\label{sympS3}
\end{equation}
Here $\imath_t:\mathbb{S}^2\rightarrow (0,\pi)\times \mathbb{S}^2$
denotes the inclusion given by $\imath_{t}(s)=(t,s)\in (0,\pi)\times
\mathbb{S}^2$. It is straightforward to show that $\Omega$ does not
depend on $t$. We will refer to the symplectic space
$\Gamma:=(\mathcal{S},\Omega)$ as the \textit{covariant phase space}
of the system.

On the other hand, we will denote the \textit{canonical phase space}
as $\Upsilon :=(\mathbf{P},\omega)$. This is the space of smooth and
symmetric\footnote{Using set theoretical language
$\mathbf{P}:=\{(Q,P)\in C^\infty(\mathbb{S}^2;\mathbb{R})\times
C^\infty(\mathbb{S}^2;\mathbb{R})\,|\, \mathcal{L}_\sigma
Q=\mathcal{L}_\sigma P=0\}\,.$} Cauchy data $(Q,P)\in\mathbf{P}$
endowed with the standard symplectic structure
\begin{eqnarray}
\omega((Q_1,P_1),(Q_2,P_2)):=\int_{\mathbb{S}^2} |\gamma|^{1/2}
(Q_2P_1-Q_1P_2)\,.
\end{eqnarray}
Given any value of $t$, the bijection
$\mathfrak{I}_t:\Upsilon\rightarrow\Gamma$, that maps every Cauchy
data $(Q,P)$ to the unique solution $\phi\in \mathcal{S}$ such that
$\phi(t,s)=Q(s)$ and $(\sin t)\dot{\phi}(t,s)=P(s)$, is a  linear
symplectomorphism $\omega=\mathfrak{I}_t^*\Omega\,.$

Elements in the linear space $\mathcal{S}$ can be expanded as\footnote{The bar denotes complex conjugation.}
\begin{equation}
\phi(t,s)=\sum_{\ell=0}^\infty\Big(a_\ell
y_\ell(t)Y_{\ell0}(s)+\overline{a_\ell
y_\ell(t)Y_{\ell0}(s)}\Big)\,, \label{expansionS3}
\end{equation}
where $Y_{\ell 0}$ denote the spherical harmonics that, in the standard spherical coordinates, have the form
$$
Y_{\ell
0}(s)=\left(\frac{2\ell+1}{4\pi}\right)^{1/2}P_\ell(\cos\theta(s))\,,
$$
in terms of Legendre polynomials $P_\ell$, and satisfy the equations
$$
\Delta_{\mathbb{S}^2}Y_{\ell 0}=-\ell(\ell+1)Y_{\ell
0}\,,\quad\mathcal{L}_\sigma Y_{\ell 0}=0\,.
$$
Notice that, modulo a global constant that we absorb in the
functions $y_\ell$, we have no other freedom in the choice of the
angular part of the modes $\phi_\ell(t,s)=y_\ell(t)Y_{\ell0}(s)$.
The coefficients $a_\ell$ must be subject, of course, to appropriate
fall-off conditions in order to guarantee the pointwise convergence
of the previous series. We also need a suitable norm in this space
to talk about convergence. Though it is possible to detail at this
stage the necessary structures and conditions we will not do so
because the final construction of the quantum Hilbert space that we
carry out is insensitive to these choices. Notice that
(\ref{expansionS3}), where only the $Y_{\ell 0}$ harmonics appear,
already takes into account the extra symmetry in the $\sigma^a$
direction.

The massless Klein-Gordon equation leads now to the following equation for the complex functions $y_\ell(t)$
\begin{equation}
\ddot{y}_\ell+(\cot t) \dot{y}_\ell+\ell(\ell+1)y_\ell=0\,.
\label{diffeqS3}
\end{equation}
We will always assume that, for each $\ell$, the real and imaginary
parts of $y_\ell$, $u_\ell$ and $v_\ell$ respectively, are two real
linearly independent solutions of (\ref{diffeqS3}). We will not make
at this point any specific choice for these functions but we will
fix their normalization in the following way. Let us substitute
first (\ref{expansionS3}) in the symplectic structure $\Omega$. We
find that
$$
\Omega(\phi_1,\phi_2)=\sin t\sum_{\ell=0}^\infty
(\bar{a}_{1\ell}a_{2\ell}-\bar{a}_{2\ell}a_{1\ell})
\big(y_\ell(t)\dot{\bar{y}}_\ell(t)-\dot{y}_\ell(t)\bar{y}_\ell(t)\big)\,.
$$
The previous expression can be simplified by first expanding $y_\ell(t)=u_\ell(t)+iv_\ell(t)$ and writing
$$
y_\ell(t)\dot{\bar{y}}_\ell(t)-\dot{y}_\ell(t)\bar{y}_\ell(t)=2i \det\left(\begin{array}{cc}
\dot{u}_\ell(t)&u_\ell(t)\\
\dot{v}_\ell(t)&v_\ell(t)
\end{array}
\right)=:2iW(t;u_\ell,v_\ell)\,.
$$
As a consequence of the fact that $y_\ell$ satisfies the differential equation (\ref{diffeqS3})
the Wronskian $W$ satisfies
$$
\dot{W}+(\cot t) W=0\Rightarrow W(t;u_\ell,v_\ell)=\frac{c_\ell}{\sin t}\,,\quad c_\ell\in \mathbb{R}\,,
$$
and hence the symplectic structure has the simple expression
\begin{equation}
\Omega(\phi_1,\phi_2)=2i\sum_{\ell=0}^\infty
c_\ell(\bar{a}_{1\ell}a_{2\ell}-\bar{a}_{2\ell}a_{1\ell})\,.
\label{symp_simp_S3}
\end{equation}
Notice that the time independence of the symplectic structure is
explicit now. In the following we will choose the pair of functions
$(u_\ell,v_\ell)$ normalized in such a way that $c_\ell=1/2$,
$\forall\,\ell$, i.e.
\begin{equation}\label{norm}
W(t;u_\ell,v_\ell)=\frac{1}{2\sin t}\,,\quad\forall\,
(u_\ell,v_\ell)\,,\quad \ell\in \mathbb{N}\cup \{0\}\,.
\end{equation}
This condition is imposed in order to ensure that the modes
$\{\phi_\ell\}_{\ell=0}^{\infty}$ define an orthogonal basis of the
one-particle Hilbert space on which we will construct the Fock space
for the quantum theory.

With the aim of characterizing the freedom in the election of the
functions $y_\ell$, let us fix a specific family
\begin{equation}
\{y_{0\ell}=u_{0\ell}+i v_{0\ell}\,|\, \ell \in \mathbb{N}\cup \{0\}\}
\label{y0}
\end{equation}
satisfying the normalization condition\footnote{Though it is
possible to choose a normalization with the opposite sign
($i\mapsto-i$), it is irrelevant as far as the unitarity issues
discussed here are concerned, and amounts to interchanging negative
and positive frequencies.} (\ref{norm})
\begin{equation}
y_{0\ell}\dot{\bar{y}}_{0\ell}-\dot{y}_{0\ell}\bar{y}_{0\ell}=\frac{i}{\sin t}\,.\label{normaliz}
\end{equation}
For any other normalized election of a family of linearly
independent functions $\{y_\ell=u_\ell+i v_\ell\}_{\ell=0}^{\infty}$
we can write (in terms of the $u_{0\ell}$ and $v_{0\ell}$)
\begin{equation}
y_\ell(t)=u_\ell (t)+iv_\ell (t)=\alpha_\ell u_{0\ell}(t)+\beta_\ell
v_{0\ell}(t)+i[\gamma_\ell u_{0\ell}(t)+\delta_\ell
v_{0\ell}(t)]\,.\label{alphabeta}
\end{equation}
The normalization that we are choosing (\ref{normaliz}) gives the
following condition for the real coefficients $\alpha_\ell$,
$\beta_\ell$, $\gamma_\ell$, and $\delta_\ell$
\begin{equation}
\alpha_\ell\delta_\ell-\beta_\ell\gamma_\ell=1\,,\quad \ell\in \mathbb{N}\cup \{0\}
\label{cond}
\end{equation}
i.e.
$$
\left(
\begin{array}{cc}
\alpha_\ell&\beta_\ell\\
\gamma_\ell&\delta_\ell
\end{array}
\right)\in SL(2;\mathbb{R})\,,\quad  \ell\in \mathbb{N}\cup \{0\}\,.
$$
As a set, $SL(2,\mathbb{R})$ is in one-to-one correspondence with
$\mathbb{S}^1\times\mathbb{R}^2$ and thus its elements can be
factorized as
\begin{equation}
SL(2,\mathbb{R})\ni\left(
\begin{array}{cc}
\alpha_\ell&\beta_\ell\\
\gamma_\ell&\delta_\ell
\end{array}
\right)=
\left(
\begin{array}{cc}
\cos\theta_\ell&-\sin\theta_\ell\\
\sin\theta_\ell&\cos\theta_\ell
\end{array}
\right)
\left(
\begin{array}{cc}
\rho_\ell&\nu_\ell\\
0&\rho_\ell^{-1}
\end{array}
\right)
\label{polar}
\end{equation}
for a unique choice of $\rho_\ell>0$, $\nu_\ell\in\mathbb{R}$,
$\theta_\ell\in[0,2\pi)$. We will show in section
\ref{FockQuantization} that the rotation part defined by the angle
$\theta_{\ell}$ plays a trivial role in the quantization of the
model. As a consequence of this we will concentrate on the other
factor involving $\rho_\ell$ and $\nu_\ell$,
\begin{equation}
y_\ell(t)=\rho_\ell u_{0\ell}(t)+(\nu_\ell+i\rho_\ell^{-1})v_{0\ell}(t)\label{y_l}\,,
\end{equation}
and choose
\begin{eqnarray}
u_{0\ell}(t)=\frac{1}{\sqrt{2}}P_\ell(\cos t),\quad v_{0\ell}(t)=\frac{1}{\sqrt{2}}Q_\ell(\cos t)\,,\quad \quad\ell\in \mathbb{N}\cup \{0\}\label{u0v0}
\end{eqnarray}
with $P_\ell$ and $Q_\ell$ denoting the first and second class
Legendre functions, respectively. As we will see in section
\ref{Invariant complex forms}, the different choices of
$\{(\rho_\ell,\nu_\ell)\,|\,\ell\in\mathbb{N}\cup \{0\} \}$ in
(\ref{y_l}) will parameterize convenient complex structures that
will allow us to construct the Fock representations for the quantum
counterpart of the system.

\subsection{Classical dynamics}

Let us consider now the classical time evolution of the system.
Given two values of the time parameter $0<t_0\leq t_1<\pi$, the
evolution from $t_0$ to $t_1$ can be viewed as a symplectomorphism
$\mathcal{T}_{(t_0,t_1)}:\Gamma\rightarrow \Gamma$ in the covariant
phase space. It is possible to write
$\mathcal{T}_{(t_0,t_1)}=\mathfrak{I}_{t_0}\circ
\mathfrak{I}^{-1}_{t_1}$ in  terms of the maps $\mathfrak{I}_t$
that, given a value of $t$, identify the space of Cauchy data
$\Upsilon$ with the covariant phase space $\Gamma$. This application
(i) takes a solution of $\mathcal{S}$, (ii) finds the Cauchy data
that this solution induces on $\iota_{t_1}(\mathbb{S}^{2})$ by
virtue of the variational principle, (iii) imposes them as initial
data on $\iota_{t_0}(\mathbb{S}^{2})$, and (iv) finally finds the
corresponding solution of $\mathcal{S}$. Explicitly, given
\begin{eqnarray}
\phi(t,s)=\sum_{\ell=0}^\infty\Big( a_\ell
y_\ell(t)Y_{\ell0}(s)+\overline{a_\ell
y_\ell(t)Y_{\ell0}(s)}\Big)\in \Gamma\,, \label{fourier}
\end{eqnarray}
the map
\begin{eqnarray}
\mathfrak{I}^{-1}_{t_1}:\Gamma\rightarrow \Upsilon\,,\quad \phi\mapsto(Q,P)=\mathfrak{I}^{-1}_{t_1}(\phi)
\end{eqnarray}
is defined by
\begin{eqnarray}
Q(s)&:=&\phi(t_1,s)=\sum_{\ell=0}^\infty \Big( a_\ell y_\ell(t_1)Y_{\ell0}(s)+\overline{a_\ell y_\ell(t_1)Y_{\ell0}(s)}\Big)\,,\label{I_inv}\\
P(s)&:=&\sin t_1\, \dot{\phi}(t_1,s)=
\sin t_1\sum_{\ell=0}^\infty \Big(a_\ell \dot{y}_\ell(t_1)Y_{\ell0}(s)+\overline{a_\ell \dot{y}_\ell(t_1)Y_{\ell0}(s)}\Big)\nonumber\,.
\end{eqnarray}
On the other hand
\begin{eqnarray}
\mathfrak{I}_{t_0}: \Upsilon\rightarrow \Gamma\,,\quad (Q,P)\mapsto  \phi=\mathfrak{I}_{t_0}(Q,P)
\end{eqnarray}
is defined,  in terms of the Fourier coefficients $a_\ell$ of $\phi$
(\ref{fourier}), by
\begin{equation}
a_\ell(t_0)=- i\sin t_0
\dot{\bar{y}}_\ell(t_0)\int_{\mathbb{S}^2}|\gamma|^{1/2} Y_{\ell
0}Q+i\bar{y}_\ell(t_0)\int_{\mathbb{S}^2}|\gamma|^{1/2} Y_{\ell
0}P\,.\label{I_direc}
\end{equation}
By using (\ref{I_inv}) and (\ref{I_direc}) we finally get
\begin{eqnarray}
(\mathcal{T}_{(t_0,t_1)}\phi)(t,s)&:=&(\mathfrak{I}_{t_0}\circ \mathfrak{I}^{-1}_{t_1}\phi)(t,s)\label{Tt0t1}\\
&=&\sum_{\ell=0}^\infty \Big(\mathfrak{a}_\ell(t_0,t_1)
y_\ell(t)Y_{\ell0}(s)+\overline{\mathfrak{a}_\ell(t_0,t_1)
y_\ell(t)Y_{\ell0}(s)}\Big)\nonumber
\end{eqnarray}
where
\begin{eqnarray}
\mathfrak{a}_\ell(t_0,t_1) &:=&-i
[\sin t_0y_\ell(t_1)\dot{\bar{y}}_\ell(t_0)-\sin t_1 \bar{y}_\ell(t_0)\dot{y}_\ell(t_1)]a_\ell
\label{timeevolution}
\\
& &-i
[\sin t_0\bar{y}_\ell(t_1)\dot{\bar{y}}_\ell(t_0)-\sin t_1\bar{y}_\ell(t_0)\dot{\bar{y}}_\ell(t_1)]\bar{a}_\ell
\,.\nonumber
\end{eqnarray}
In the next sections we will try to find out if this classical evolution can be unitarily implemented in a Fock quantization of the system.

\section{Fock quantization}{\label{FockQuantization}}

In the passage to the quantum theory we have to introduce a Hilbert
space for our system that we will write as
$\bigotimes_{i=0}^N\mathcal{F}_i$. The Hilbert spaces
$\mathcal{F}_i$ will be used to describe the gravitational modes
$(i=0)$ and the massless scalar fields $(i\in\mathbb{N})$. These
will be taken to be symmetric Fock spaces built from appropriate one
particle Hilbert spaces. As they are all isomorphic, and all the
massless scalars satisfy the same equation, the same construction
will be valid for all of them so we will omit the $i$ index in the
following. Here we will follow the quantization steps discussed in
section 2.3 of reference \cite{Wald} in order to define a suitable
separable physical Hilbert space for the quantum theory, as well as
irreducible representations for the canonical commutation relations.
As expected for scalar fields in non-stationary curved background
space-times, the Fock representation obtained in this way is highly
non-unique.

In order to define the one-particle Hilbert space used to build the
Fock space $\mathcal{F}$, let
$\mathcal{S}_{\mathbb{C}}:=\mathbb{C}\otimes \mathcal{S}$ denote the
$\mathbb{C}$-vector space obtained by the complexification of the
solution space $\mathcal{S}$ introduced above (\ref{solspace}). The
elements of $\mathcal{S}_{\mathbb{C}}$ are ordered pairs of objects
$(\phi_1,\phi_2)\in\mathcal{S}\times\mathcal{S}$ that we will write
in the form\footnote{Here $i\in\mathbb{C}$ denotes the imaginary
unit.} $\Phi:=\phi_1+i\phi_2$ with the natural definition for their
sum. Multiplication by complex scalars
$\mathbb{C}\ni\lambda=\lambda_1+i\lambda_2$,
$\lambda_1,\lambda_2\in\mathbb{R}$, is defined as
$$
\lambda\Phi:=(\lambda_1\phi_1-\lambda_2\phi_2)+i(\lambda_2\phi_1+\lambda_1\phi_2).
$$
We also introduce the conjugation
$\bar{\phantom{z}}:\mathcal{S}_{\mathbb{C}}\rightarrow
\mathcal{S}_{\mathbb{C}}:(\phi_1+i\phi_2)\mapsto(\phi_1-i\phi_2)$.
Vectors in $\mathcal{S}_{\mathbb{C}}$ can be expanded with the help
of the basis $\{\phi_\ell:=y_\ell Y_{\ell 0},
\bar{\phi}_\ell:=\overline{y_\ell Y_{\ell 0}}\}_{\ell=0}^\infty$
introduced above as
$$\Phi=\sum_{\ell=0}^\infty\Big(a_\ell y_\ell
Y_{\ell 0}+b_\ell\overline{y_\ell Y_{\ell 0}}\,\Big)$$ with
$a_\ell,b_\ell\in\mathbb{C}$. The symplectic structure
(\ref{sympS3}) defined on $\mathcal{S}$ can be extended in a linear
way to $\mathcal{S}_{\mathbb{C}}$ as
$$
\Omega_{\mathbb{C}}(\Phi_1,\Phi_2):=i\sum_{\ell=0}^\infty\Big(b_{1\ell}a_{2\ell}-b_{2\ell}a_{1\ell}\Big)\,.
$$
For each pair $\Phi_1,\Phi_2\in\mathcal{S}_{\mathbb{C}}$ the mapping
\begin{equation}
\langle\cdot|\cdot\rangle:\mathcal{S}_{\mathbb{C}}\times
\mathcal{S}_{\mathbb{C}}\rightarrow\mathbb{C},\quad(\Phi_1,\Phi_2)\mapsto\langle\Phi_1|\Phi_2\rangle:=-i\Omega_{\mathbb{C}}(\bar{\Phi}_1,\Phi_2)
\label{escprod}
\end{equation}
is antilinear in the first argument and linear in the second. It is
not an inner product because it is not positive-definite. There are,
however, linear subspaces of $\mathcal{S}_{\mathbb{C}}$ where
$\langle\cdot|\cdot\rangle$ is positive definite (and, hence,
defines an inner product). Let us consider, in particular, the
Lagrangian subspace
\begin{equation}
\mathcal{P}:=\big\{\Phi\in\mathcal{S}_{\mathbb{C}}\,|\,\Phi=\sum_{\ell=0}^\infty
a_\ell\phi_\ell\big\}\,. \label{P}
\end{equation}
Here the restriction $\langle\cdot|\cdot\rangle|_\mathcal{P}$
defines an inner product given by
\begin{equation}\label{inprod}
\langle \Phi_1|\Phi_2\rangle=\sum_{\ell=0}^\infty
\bar{a}_{1\ell}a_{2\ell}\,,\quad \Phi_{1},\Phi_{2}\in\mathcal{P}\,.
\end{equation}
The one particle Hilbert space $\mathcal{H}_\mathcal{P}$ is then the
Cauchy completion of
$(\mathcal{P},\langle\cdot|\cdot\rangle|_\mathcal{P})$ w.r.t. the
norm defined by the inner product. Notice that the set
$\{\phi_\ell=y_\ell Y_{\ell 0}\,|\,\ell\in \mathbb{N}\cup\{0\}\}$
becomes an orthonormal basis of $\mathcal{H}_\mathcal{P}$ satisfying
$\langle \phi_{\ell_1}\,|\,\phi_{\ell_2
}\rangle=\delta(\ell_1,\ell_2)$. Finally, the quantum Hilbert space\footnote{At variance with the $\mathbb{T}^3$ case where some constraints must be taken into account we do not have any in this case.} is given by the symmetric Fock space
$$
\mathcal{F}_s(\mathcal{H}_\mathcal{P})=\displaystyle
\bigoplus_{n=0}^{\infty}\mathcal{H}_{\mathcal{P}}^{\otimes_s
n},
$$
where $\mathcal{H}_{\mathcal{P}}^{\otimes_s
0}:=\mathbb{C}$, and
$\mathcal{H}_{\mathcal{P}}^{\otimes_s
n}$ denotes the subespace of
$\mathcal{H}_{\mathcal{P}}^{\otimes
n}=\otimes_{k=1}^n\mathcal{H}_\mathcal{P}$ spanned by symmetric
tensor products of $n$ vectors in $\mathcal{H}_\mathcal{P}$ (these are referred to as $n$-particle subspaces). Associated to the modes $\phi_\ell\in\mathcal{H}_\mathcal{P}$ we
have the corresponding annihilation $\hat{a}_\ell$ and creation
operators $\hat{a}_\ell^\dagger$, with non-vanishing commutation relations given by $[\hat{a}_{\ell_1},\hat{a}_{\ell_2}^\dagger]=\delta(\ell_1,\ell_2)$. As usual, we will denote as
$|0\rangle$ the Fock vacuum $1\oplus 0\oplus 0\oplus
\cdots \in \mathcal{F}_s(\mathcal{H}_\mathcal{P})$ whose only
nonzero component is $1\in\mathbb{C}$ and we will use a subindex $\mathcal{P}$ whenever we have to emphasize the dependence of these objects on the subspace $\mathcal{P}$.
The Fock vacuum  $|0\rangle$ is in the domain of all finite products of creation and annihilation operators and the vectors
$$
| ^1\! n_{\ell_1}\, ^2\!n_{\ell_2}\,\cdots \,^k\! n_{\ell_k}\rangle :=\frac{1}{\sqrt{ ^1n!\,^2n!\,\cdots\,^kn!}} (\hat{a}_{\ell_1}^{\dagger})^{ ^1 \! n}(\hat{a}_{\ell_1}^{\dagger})^{ ^2\! n} \cdots (\hat{a}_{\ell_k}^{\dagger})^{ ^k \! n} |0\rangle\in \mathcal{F}_s(\mathcal{H}_\mathcal{P}) \,,
$$
where   $k\in \mathbb{N}\cup \{0\}$,  $(\,\!^1\!n,\, ^2\!n,\dots,\, ^k\!n)\in \mathbb{N}^k$, and $\ell_i\neq \ell_j$ for $i\neq j$, provide a basis of $\mathcal{F}_s(\mathcal{H}_\mathcal{P})$. The basis vectors are normalized according to
\begin{eqnarray*}
\langle \, ^1\! n_{\ell_1}\,\cdots \,^k\! n_{\ell_k} \,|\,^1\! m_{\ell'_1}\,\cdots \,^r\! m_{\ell'_r} \rangle =
\delta(k, r)     \sum_{\pi\in S_k} \delta( ^1\! n, ^{\pi(1)}\! m) \cdots  \delta( ^k\!n, ^{\pi(k)}\! m)
\delta(\ell_1,\ell'_{\pi(1)}) \cdots  \delta( \ell_k,\ell'_{\pi(k)})\,,
\end{eqnarray*}
where $S_k$ denotes the set of permutations $\pi$ of the $k$ symbols $\{1,2,\dots,k\}$. Also, they satisfy
$$
\hat{a}^\dagger_\ell |n_\ell\rangle=\sqrt{n+1}\,|(n+1)_\ell\rangle\,,\quad \hat{a}_\ell |n_\ell\rangle=\sqrt{n}\,|(n-1)_\ell\rangle\,.
$$
Notice that, using the notation introduced above, the modes  $\phi_\ell$  of the one particle Hilbert space $\mathcal{H}_\mathcal{P}$ can now be considered as one-particle states that we will denote as $|1_\ell\rangle:=a^\dagger_\ell|0\rangle\in \mathcal{F}_s(\mathcal{H}_\mathcal{P})$.

\subsection{Complex structures}{\label{Complex forms}}

The previous construction for the one-particle Hilbert space is
based on a non-unique choice (\ref{P}) for the subspace
$\mathcal{P}$ of what are usually called ``positive frequency"
solutions to the field equations. Since we are dealing with
non-stationary space-times, it is not possible to select a natural
subspace $\mathcal{P}$ by invoking a time translation symmetry.
Furthermore, the deparameterization procedure does not provide extra
constraints \cite{BarberoG.:2007} that would generate residual
symmetries useful to define a preferred choice of $\mathcal{P}$.
This fact manifests itself as an ambiguity in the formulation of the quantum
theory, because different choices of $\mathcal{P}$ generally yield
unitarily inequivalent Fock representations \cite{Wald}. We will
show here that every possible choice of the subspace $\mathcal{P}$ defined in (\ref{P}) is in
correspondence with a $SO(3)$-invariant complex structure on the
solution space $\mathcal{S}$, postponing to section
\ref{uniqueness} a discussion of the uniqueness of the
representation.

An equivalent way to deal with  the splitting
$\mathcal{S}_{\mathbb{C}}=\mathcal{P}\oplus \bar{\mathcal{P}}$ is to
introduce a complex structure $J:\mathcal{S}_{\mathbb{C}}\rightarrow
\mathcal{S}_{\mathbb{C}}$, and define $\mathcal{P}$ and
$\bar{\mathcal{P}}$ as the eigenspaces associated to the eigenvalues
$+i$ and $-i$, respectively. The complex structure must satisfy the
following conditions
\begin{itemize}
\item[J1)] $J$ is a $\mathbb{C}$-linear map $J: \mathcal{S}_\mathbb{C}\rightarrow \mathcal{S}_\mathbb{C}$ satisfying $J^2=-\mathrm{Id}_{\mathcal{S}_\mathbb{C}}$.
\item[J2)] $J$ induces a $\mathbb{R}$-linear map $\mathcal{S}\rightarrow \mathcal{S}$ i.e. $\overline{J \Phi}=J\bar{\Phi}$ for all $\Phi\in \mathcal{S}_\mathbb{C}$.
\item[J3)] The sesquilinear form (\ref{escprod}) restricted to the subspace corresponding to the $i$ eigenvalue of $J$ (that we denote as $\mathcal{P}$) defines an inner product.
\end{itemize}
In practice this complex structure is defined once a choice of modes
like the one introduced above is given. For example, if we consider
the family $\{y_{0\ell}\}_{\ell=0}^{\infty}$ given by (\ref{y0}) and
(\ref{u0v0}), the set of functions $\{\phi_{0\ell}=y_{0\ell}
Y_{\ell}\}_{\ell=0}^{\infty}$ allows us to define a complex
structure by
$$
J_0\phi_{0\ell}:=i\phi_{0\ell}\,,\quad
J_0\bar{\phi}_{0\ell}:=-i\bar{\phi}_{0\ell}\,.
$$
We will denote the vector spaces generated by $\phi_{0\ell}$ and
$\bar{\phi}_{0\ell}$ as $\mathcal{P}_0$ and $\bar{\mathcal{P}}_0$
respectively. In principle a different choice for
$\{y_{\ell}\}_{\ell=0}^{\infty}$ would give rise to a different
complex structure. However this is not always the case. For example,
if we obtain $y_\ell$ from $y_{0\ell}$ by the rotation appearing in
the decomposition (\ref{polar}) of the $SL(2,{\mathbb{R}})$
matrices discussed above
$$
y_\ell=u_\ell+iv_\ell=\cos\theta_\ell u_{0\ell}-\sin\theta_\ell v_{0\ell}+i(\sin\theta_\ell u_{0\ell}+\cos \theta_\ell v_{0\ell} )=e^{i\theta_\ell} y_{0\ell}
$$
the set $\{\phi_{\ell}=y_\ell Y_{\ell 0}\}_{\ell=0}^{\infty}$
defines a complex structure $J$ through
$$
J\phi_{\ell }:=i\phi_{\ell}\,,\quad J\bar{\phi}_{\ell}:=-i\bar{\phi}_{\ell}\,.
$$
Now it is straightforward to see that $J\phi_{\ell }=i\phi_{\ell }
\Leftrightarrow J
e^{i\theta_\ell}\phi_{0\ell}=ie^{i\theta_\ell}\phi_{0\ell }$ and
$\mathbb{C}$-linearity implies $J \phi_{0\ell}=i\phi_{0\ell}$ i.e.
$J=J_0$.

Given the decomposition $\mathcal{S}_{\mathbb{C}}=\mathcal{P}_0\oplus\bar{\mathcal{P}}_0$
there are two antilinear maps that connect the spaces $\mathcal{P}_0$ and $\bar{\mathcal{P}}_0$ that we denote (in a slight notational abuse) with the same symbol\, $\bar{\phantom{z}}:\mathcal{P}_0\rightarrow\bar{\mathcal{P}}_0:\psi_1 \mapsto \bar{\psi}_1$ and
$\bar{\phantom{z}}:\bar{\mathcal{P}}_0\rightarrow \mathcal{P}_0:\psi_2\mapsto \bar{\psi}_2$. Each one of these maps is the inverse of the other and their composition is the identity  for every element of $\mathcal{P}_0$ or $\bar{\mathcal{P}}_0$ (i.e. $\bar{\bar{\psi}}=\psi$). With their help we can write the conjugation $\bar{\phantom{z}}:\mathcal{S}_{\mathbb{C}}\rightarrow\mathcal{S}_{\mathbb{C}}$ according to
$$
\Psi=
\left(\begin{array}{c}\psi_1\\\psi_2\end{array}\right)\mapsto \bar{\Psi}:=\left(\begin{array}{c}\bar{\psi}_2\\\bar{\psi}_1\end{array}\right)
$$
with $\psi_1\in{\mathcal{P}}_0$ and $\psi_2\in{\bar{\mathcal{P}}}_0$.
The elements in the original (real) solution space $\mathcal{S}$ can
be easily characterized by using the previous conjugation as those of the form
$$
\Phi=\left(\begin{array}{c}\phi\\\bar{\phi}\end{array}\right)
$$
or, alternatively, as the real linear subspace of
$\mathcal{S}_{\mathbb{C}}$ given by
$\mathcal{S}=\{\Phi\in\mathcal{S}_{\mathbb{C}}\,|\,
\Phi=\bar{\Phi}\}$.

Let us characterize now the complex structures in
$\mathcal{S}_{\mathbb{C}}$ with the help of the fixed decomposition
introduced above
$\mathcal{S}_{\mathbb{C}}=\mathcal{P}_0\oplus\bar{\mathcal{P}}_0$.
In particular, every  $\mathbb{C}$-linear map
$J:\mathcal{S}_{\mathbb{C}}\rightarrow\mathcal{S}_{\mathbb{C}}$ can
be written in the form
$$
J=\left(\begin{array}{cc}J_{11}&J_{12}\\J_{21}&J_{22}\end{array}\right),
$$
where the maps $J_{ab}:\mathcal{P}_b\rightarrow \mathcal{P}_a$ are
$\mathbb{C}$-linear for $a,b\in\{1,2\}$, and we have introduced the
convenient notation $\mathcal{P}_1:=\mathcal{P}_0$ and
$\mathcal{P}_2:=\bar{\mathcal{P}}_0$. The necessary and sufficient
condition to guarantee that the restriction of $J$ to $\mathcal{S}$
is $\mathbb{R}$-linear is $J\Phi=\overline{J\Phi}$ for every
$\Phi\in\mathcal{S}$, or equivalently
\begin{eqnarray*}
J_{11}\phi=\overline{J_{22}\bar{\phi}}\,,\quad
J_{21}\phi=\overline{J_{12}\bar{\phi}}\,,
\end{eqnarray*}
i.e.
\begin{equation}
J_{22}=\bar{J}_{11}\,,\quad J_{12}=\bar{J}_{21}\,,\label{JbarJ}
\end{equation}
where we have used the notation
$\bar{A}\phi:=\overline{A\bar{\phi}}$ to denote the
$\mathbb{C}$-linear map $\bar{A}:P_b\rightarrow P_a$ ($a\neq b$)
obtained from the $\mathbb{C}$-linear map $A:P_a\rightarrow P_b$.
Finally the condition $J^2=-\mathrm{Id}_{\mathcal{S}_{\mathbb{C}}}$
requires that
$$
J^2_{11}+\bar{J}_{21}J_{21}=-\mathrm{Id}_1\,,\quad
J_{21}J_{11}+\bar{J}_{11}J_{21}=0\,.
$$
We will see in the next subsection how the symmetries of the problem
help us fix the form of the $J_{ab}$.

\subsection{Invariant complex structures}{\label{Invariant complex forms}}

Here we want to characterize those complex structures in the
solution space $\mathcal{S}^{KG}$ of the field equation
$\mathring{g}^{ab}\mathring{\nabla}_a\mathring{\nabla}_b\phi=0$,
invariant under the symmetries of $\mathbb{S}^2$ --the spatial
manifold in our $(2+1)$-dimensional description-- \textit{without
imposing the condition} $\mathcal{L}_\sigma\phi=0$. As we will show,
once this is done it is straightforward to restrict them to the
solution space $\mathcal{S}$. To this end let us consider the
complexified solution space
$\mathcal{S}^{KG}_{\mathbb{C}}=\mathcal{P}^{KG}_0\oplus\bar{\mathcal{P}}^{KG}_0$
where \begin{eqnarray*}
\mathcal{P}^{KG}_1&:=&\mathcal{P}^{KG}_0=\mathrm{span}\{y_{0\ell}
Y_{\ell m}\,|\, \ell\in \mathbb{N}\cup\{0\},\,
m\in\{-\ell,\dots,\ell\}\}\,,\\
\mathcal{P}^{KG}_2&:=&\bar{\mathcal{P}}^{KG}_0=\mathrm{span}\{\bar{y}_{0\ell}
Y_{\ell m} \,|\, \ell\in \mathbb{N}\cup\{0\},\,
m\in\{-\ell,\dots,\ell\}\}\,.
\end{eqnarray*}
Here $Y_{\ell m}$ are the usual spherical harmonics on
$\mathbb{S}^2$.

The elements $\phi_a\in \mathcal{P}^{KG}_a$, $a=1,2$, are complex
functions $\phi_a(t,s)$ defined on $(0,\pi)\times \mathbb{S}^2$.
There is a natural representation $D_a$  of $SO(3)$ in
$\mathcal{P}^{KG}_a$ defined by
$(D_a(g)\phi)(t,s)=\phi(t,g^{-1}\cdot s)$ where $g^{-1}\cdot s$
denotes the action of the rotation $g^{-1}\in SO(3)$ on the point
$s\in \mathbb{S}^2$.  Then the natural representation of $SO(3)$ in
$\mathcal{S}^{KG}_{\mathbb{C}}=\mathcal{P}^{KG}_1\oplus\mathcal{P}^{KG}_2$
can be written in matrix form as
$$
D(g)=\left(\begin{array}{cc}
D_1(g)&0\\0&D_2(g)
\end{array}\right),\quad g\in SO(3)\,,
$$
in terms of the representations $(D_a,\mathcal{P}^{KG}_a)$. The
invariance of a $\mathbb{C}$-linear map $J$ under the action of the
group $SO(3)$ implies
$$
D(g)J=J D(g)\Leftrightarrow \left(\begin{array}{cc}
J_{11}D_1(g)&J_{12}D_2(g)\\J_{21}D_1(g)&J_{22}D_2(g)
\end{array}\right)=\left(\begin{array}{cc}
D_1(g)J_{11}&D_1(g)J_{12}\\D_2(g)J_{21}&D_2(g)J_{22}
\end{array}\right),\quad \forall g\in SO(3)\,.
$$
It is convenient now to expand the vector spaces $\mathcal{P}^{KG}_a$ as
\begin{eqnarray}
\mathcal{P}^{KG}_a=\bigoplus_{\ell=0}^{\infty}
\mathcal{P}^\ell_a\,,\quad a=1,2\,,
\end{eqnarray}
with
\begin{eqnarray*}
\mathcal{P}^\ell_1&:=&\mathrm{span}\{y_{0\ell}\}\otimes\mathrm{span}\{Y_{\ell
m}\,|\, m\in\{-\ell,\dots,\ell\}\}\,, \\
\mathcal{P}_2^\ell&:=&\mathrm{span}\{\bar{y}_{0\ell}\}\otimes\mathrm{span}\{Y_{\ell
m}\,|\, m\in\{-\ell,\dots,\ell\}\}\,.
\end{eqnarray*}
This is useful because the operators $D_a(g)$ can be written as $D_a=\bigoplus_{\ell=0}^\infty D_a^\ell$, where each of the $(\mathcal{P}_a^\ell,D_a^\ell)$ are irreducible representations.

Denoting as $\Pi_a^\ell$ the projectors on the linear spaces $\mathcal{P}_a^\ell$ we can write the linear mappings
$J_{ab}$ as
$$
J_{ab}^{\ell_1\ell_2}:=\Pi_a^{\ell_1}J_{ab}\Pi_b^{\ell_2}:\mathcal{P}_b^{\ell_2}\rightarrow \mathcal{P}_a^{\ell_1}.
$$
We use now Schur's lemma\footnote{\textit{Schur lemma:} Let $D_1(g)$ and $D_2(g)$ be two finite dimensional, irreducible representations of the group $G$ in the complex finite-dimensional linear spaces $V_1$ and $V_2$. Let us suppose that a linear operator $L:V_1\rightarrow V_2$ `commutes' with these representations (i.e. $D_2(g)L=LD_1(g)$, $\forall g\in G$). Then either $L$ is zero or it is invertible. In this last case both representations are equivalent and $L$ is uniquely determined modulo a multiplicative constant.} that directly implies that $J_{ab}^{\ell_1\ell_2}=0$ whenever $\ell_1\neq \ell_2$, $J_{aa}^{\ell\ell}=\jmath_{aa}^\ell I_{aa}^\ell$
where $\jmath_{aa}^\ell\in\mathbb{C}$, ($I_{aa}^\ell$ denotes the identity on $\mathcal{P}_a^\ell$) and $\jmath_{22}^\ell=\bar{\jmath}^\ell_{11}$ as a consequence of (\ref{JbarJ}). Also
$$
J_{12}^{\ell\ell}(\bar{y}_{0\ell}\otimes v)=\jmath_{12}^\ell
y_{0\ell}\otimes v,\quad J_{21}^{\ell\ell}(y_{0\ell}\otimes
v)=\jmath_{21}^\ell \bar{y}_{0\ell}\otimes v,\quad
\jmath_{12}^\ell,\,\jmath_{21}^\ell\in\mathbb{C}$$ with
$\jmath_{12}^\ell=\bar{\jmath}_{21}^\ell$ again as a consequence of
(\ref{JbarJ}). In conclusion the general form of the mapping $J$ is
given by
$$
J=\bigoplus_{\ell=0}^{\infty}\left(
\begin{array}{cc}
\jmath_{11}^\ell I_{11}^\ell&\jmath_{12}^\ell I_{12}^\ell\\
\bar{\jmath}_{12}^\ell I_{21}^\ell&\bar{\jmath}_{11}^\ell
I_{22}^\ell
\end{array}
\right),
$$
where $I_{aa}^\ell$ denotes the identity operator in $\mathcal{P}_a^\ell$ and the linear operators $I_{ab}^\ell:\mathcal{P}_b^\ell\rightarrow\mathcal{P}_a^\ell$ act according to $I_{12}^\ell(\bar{y}_{0\ell}\otimes v)=y_{0\ell}\otimes v$ and $I_{21}^\ell(y_{0\ell}\otimes v)=\bar{y}_{0\ell}\otimes v$.

The condition $J^2=-\mathrm{Id}_{\mathcal{S}_{\mathbb{C}}^{KG}}$
defining $J$ as a complex structure gives finally the following
restriction on $\jmath_{11}^\ell$ and $\jmath_{12}^\ell$
\begin{equation}
|\jmath_{11 }^\ell|\,^2-|\jmath_{12}^\ell|\,^2=1,\quad \jmath_{11
}^\ell\in i \mathbb{R}\smallsetminus\{0\}\,,\quad
\jmath_{12}^\ell\in \mathbb{C}\,. \label{conds}
\end{equation}
Several comments are in order now. First of all as we can see, on
each subspace $\mathcal{P}^\ell_1\oplus\mathcal{P}^\ell_2$ the
complex structure is completely fixed by a pair of complex
parameters $(\jmath_{11 }^\ell,\jmath_{12}^\ell)$ subject to the
conditions (\ref{conds}); the remaining freedom is then
parameterized by two real numbers. This is what we have found before
by explicitly considering the solution space and the choice of the
families of functions $u_\ell$ and $v_\ell$. It is straightforward
to check that the complex structures naturally defined by these
families of functions are in fact $SO(3)$ invariant. The previous
argument then shows that they exhaust, in fact, all the
possibilities. The choice $\jmath_{11}^\ell\in i\mathbb{R}_+$ is
equivalent to the normalization for the Wronskian of $u_\ell$ and
$v_\ell$ introduced above in equation (\ref{norm}) and guarantees
that the condition J3 in section \ref{Complex forms} is satisfied.
Changing the sign in the Wronskian corresponds to taking
$\jmath_{11}^\ell\in i\mathbb{R}_-$.

The previous considerations apply to solutions of the
Klein-Gordon equation without imposing the additional axial
symmetry. This can be trivially taken into account at this point by
realizing that it suffices to restrict ourselves to the
one-dimensional subspaces (for each value of $\ell$) spanned by the
spherical harmonics $Y_{\ell0}$.

Finally we give here the formulas that relate the parameters
$\rho_\ell$ and $\nu_\ell$ to the definition of the invariant
complex structure discussed in this section. Once a fiducial basis
$\phi_{0\ell}=y_{0\ell}Y_{\ell0}$ is chosen (\ref{y0}) any other
complex structure defined by a different basis --satisfying the
normalization condition (\ref{normaliz})-- can be written in terms
of $\phi_{0\ell}$, by using (\ref{alphabeta}) and (\ref{polar}), as
\begin{eqnarray}
J\left(\begin{array}{c}\phi_{0\ell}\\\bar{\phi}_{0\ell}\end{array}\right)=\left(\begin{array}{cc} \jmath_{11}^\ell I_{11}^\ell & \jmath_{12}^\ell \, I_{12}^\ell\\
 \bar{\jmath}_{12}^\ell \, I_{21}^\ell& \bar{\jmath}_{11}^\ell\,
 I_{22}^\ell\end{array}\right)\left(\begin{array}{c}\phi_{0\ell}\\\bar{\phi}_{0\ell}\end{array}\right),
\label{JJ0}
\end{eqnarray}
where
\begin{eqnarray}
\jmath_{11}^\ell&=&\frac{i}{2}(\alpha^2_\ell+\beta^2_\ell+\gamma^2_\ell+\delta^2_\ell)
=\frac{i}{2}(\nu_\ell^2+\rho^{-2}_\ell+\rho^2_\ell)\,,\label{j1}  \\
\jmath_{12}^\ell&=&
-(\alpha_\ell\beta_\ell+\gamma_\ell\delta_\ell)+\frac{i}{2}(\beta^2_\ell+\delta^2_\ell-\alpha^2_\ell-\gamma^2_\ell)
=-\rho_\ell\nu_\ell+\frac{i}{2}(\nu_\ell^2+\rho^{-2}_\ell-\rho^2_\ell)\,. \label{j2}
\end{eqnarray}
Notice that, as expected, the complex structures defined by
(\ref{j1}) and (\ref{j2}) do not depend on the parameters
$\theta_\ell\in[0,2\pi)$ appearing in (\ref{polar}) but only on the
pairs $(\rho_\ell,\nu_\ell)\in (0,\infty)\times \mathbb{R}$. Notice
also that these last formulas relate the invariant complex
structures described here with the ones obtained in section
\ref{Complex forms} by studying the mode decomposition in the
solution space.

\section{Unitarity of the quantum time evolution and uniqueness of the Fock representation}{\label{UnitarityEvolution}}

We discuss in this section the unitarity of the quantum evolution
for the classical system described in section \ref{PhaseSpace}
corresponding to the reduced phase space of the Gowdy models coupled
to massless scalar fields with  $\mathbb{S}^1\times \mathbb{S}^2$
and $\mathbb{S}^3$ spatial topologies. We also study the uniqueness
(after re-scaling of the field) of the Fock representation under the requirement, on the complex structures, of $SO(3)$ invariance and unitarity of the dynamics.

It is well known \cite{Shale} that not every linear symplectic
transformation $\mathcal{T}$ defined on the infinite dimensional
symplectic linear space $\Gamma$ can be unitarily implemented in a
Fock quantization  of the system.  Let
$\mathcal{T}:\Gamma\rightarrow \Gamma$ be a \textit{continuous}
linear symplectic  transformation. Given any point
\begin{eqnarray}
\phi=\sum_{\ell=0}^\infty \Big(\, a_\ell y_\ell
Y_{\ell0}+\overline{ a_\ell y_\ell Y_{\ell0}}\,\Big)\in \Gamma\,,
\end{eqnarray}
 the action of $\mathcal{T}$  can be written in the form
$$
\mathcal{T}\phi=\sum_{\ell_1=0}^\infty \Big( \mathfrak{a}_{\ell_1}(a,\bar{a})y_{\ell_1}Y_{\ell_1 0} + \overline{\mathfrak{a}_{\ell_1}(a,\bar{a})y_{\ell_1} Y_{\ell_1 0}} \Big)\,,
$$
where the complex coefficients
$$
\mathfrak{a}_{\ell_1}(a,\bar{a})=\sum_{\ell_2=0}^\infty \Big(\alpha(\ell_1,\ell_2) a_{\ell_2}+\beta(\ell_1,\ell_2)\bar{a}_{\ell_2}\Big)\,
$$
must satisfy certain conditions to ensure the continuity. $\mathcal{T}$ is
implementable in the quantum theory as a unitary operator, i.e.
there exists a unitary operator
$\hat{U}_{\mathcal{T}}:\mathcal{F}_s(\mathcal{H}_\mathcal{P})\rightarrow
\mathcal{F}_s(\mathcal{H}_\mathcal{P})$ such that
$$
\hat{U}_\mathcal{T}^{-1}\hat{a}_{\ell_1}\, \hat{U}_\mathcal{T}=
\sum_{\ell_2=0}^\infty \Big(\alpha(\ell_1,\ell_2) \hat{a}_{\ell_2}+\beta(\ell_1,\ell_2)\hat{a}^\dag_{\ell_2}\Big)\,,
$$
if and only if  $J_{\mathcal{P}}-\mathcal{T}^{-1}\circ J_{\mathcal{P}}\circ \mathcal{T}$ is Hilbert-Schmidt (here $J_{\mathcal{P}}$ is the complex structure associated to the $\mathcal{P}$ subspace) \cite{Shale,Wald}. Equivalently this can be expressed as
$$
\sum_{\ell_1=0}^\infty\sum_{\ell_2=0}^\infty|\beta(\ell_1,\ell_2)|^2<\infty\,.
$$
This condition for the unitary implementability of the  symplectic transformation
$\mathcal{T}_{(t_0,t_1)}$, that defines the time evolution on $\Gamma$, can be written from (\ref{Tt0t1}) and (\ref{timeevolution}) as
\begin{eqnarray}
\sum_{ \ell=0}^{\infty}|\beta_\ell(t_0,t_1|y_\ell)|^2 =
\sum_{\ell=0}^{\infty}|\sin t_0 y_\ell(t_1)\dot{y}_\ell(t_0)-\sin
t_1
y_\ell(t_0)\dot{y}_\ell(t_1)|^2<\infty,\,\forall\,t_0,t_1\in(0,\pi)\,,
\label{(A)}
\end{eqnarray}
where $\beta_\ell(t_0,t_1|y_\ell):=\sin t_0
y_\ell(t_1)\dot{y}_\ell(t_0)-\sin t_1 y_\ell(t_0)\dot{y}_\ell(t_1)$.
At this point we have to study the convergence of the previous
series. To this end let us consider the imaginary part of the
coefficients $\beta_\ell$; by using the expression (\ref{y_l}) for
$y_\ell$ it is possible to identify the dependence of
$\mathrm{Im}(\beta_\ell(t_0,t_1|y_\ell))$ on the choice of complex
structure --parameterized by $(\rho_\ell,\nu_\ell)$. This is given
by
\begin{eqnarray}
\mathrm{Im}(\beta_\ell(t_0,t_1|y_\ell))&=&
A_\ell(t_0,t_1)+2\rho^{-1}_\ell\nu_\ell B_\ell(t_0,t_1)\,,
\end{eqnarray}
where
\begin{eqnarray*}
A_\ell(t_0,t_1)&:=&\sin t_0[u_{0\ell}(t_1)\dot{v}_{0\ell}(t_0)+\dot{u}_{0\ell}(t_0)v_{0\ell}(t_1)]-
\sin t_1[u_{0\ell}(t_0)\dot{v}_{0\ell}(t_1)+v_{0\ell}(t_0)\dot{u}_{0\ell}(t_1)]\,,
\\
B_\ell(t_0,t_1)&:=&\sin t_0v_{0\ell}(t_1)\dot{v}_{0\ell}(t_0)-\sin t_1 v_{0\ell}(t_0)\dot{v}_{0\ell}(t_1)\,.
\end{eqnarray*}
The explicit form of $A_\ell$ and $B_\ell$, derived in a straightforward way from (\ref{u0v0}), is
\begin{eqnarray}
A_\ell(t_0,t_1)
&=&-\frac{\ell+1}{2}\Big(
P_{\ell+1}(\cos t_1)Q_\ell(\cos t_0)-P_{\ell+1}(\cos t_0)Q_\ell(\cos t_1)\\
&&\hspace{19mm}+P_\ell(\cos t_1)[(\cos t_0-\cos t_1)Q_\ell(\cos t_0)-Q_{\ell+1}(\cos t_0)]\nonumber\\
&&\hspace{19mm}+P_\ell(\cos t_0)[(\cos t_0-\cos t_1)Q_\ell(\cos t_1)-Q_{\ell+1}(\cos t_1)]\Big)\,,\nonumber
\\
B_\ell(t_0,t_1)
&=&\frac{\ell+1}{2}\Big(Q_\ell(\cos t_1)Q_{\ell+1}(\cos t_0)\nonumber\\
& & \hspace{15mm}-[(\cos t_0-\cos t_1)Q_\ell(\cos t_1)+Q_{\ell+1}(\cos t_1)]Q_\ell(\cos t_0)\Big)\,.\nonumber
\end{eqnarray}
By using the following asymptotic expansions for the first and second class Legendre functions ($\varepsilon<t<\pi-\varepsilon$, $\varepsilon>0$) \cite{Abramowitz}
\begin{eqnarray}
&& P_\ell(\cos t)=\frac{\Gamma(\ell+1)}{\Gamma(\ell+3/2)}\sqrt{\frac{2}{\pi\sin t}}\cos[(\ell+1/2)t-\pi/4]+O(\ell^{-1})\,,\label{asintoticoQP}\\
&& Q_\ell(\cos t)=\frac{\Gamma(\ell+1)}{\Gamma(\ell+3/2)}\sqrt{\frac{\pi}{2\sin t}}\cos[(\ell+1/2)t+\pi/4]+O(\ell^{-1})\,,\nonumber
\label{asympt}
\end{eqnarray}
we find that, for $\ell\rightarrow\infty$,
\begin{eqnarray*}
\mathrm{Im}(\beta_\ell(t_0,t_1|y_\ell))& \sim & -\frac{1}{2}\frac{\sin t_0-\sin t_1}{\sqrt{\sin t_0\sin t_1}}\sin[(\ell+1/2)(t_0+t_1)]\nonumber\\
&&-\frac{\pi \nu_\ell\rho_\ell^{-1}}{2\sqrt{\sin t_0 \sin t_1}}\Big(\sin t_0 \cos[(\ell+1/2)t_1+\pi/4]\sin[(\ell+1/2)t_0+\pi/4]\nonumber\\
&&\hspace{3cm}-\sin t_1
\cos[(\ell+1/2)t_0+\pi/4]\sin[(\ell+1/2)t_1+\pi/4]\Big)\,.\nonumber
\end{eqnarray*}
The asymptotic behavior of $\mathrm{Im}(\beta_\ell(t_0,t_1|y_\ell))$
leads us to conclude that irrespective of the choice of
$(\rho_\ell,\nu_\ell)$ we have that
$\mathrm{Im}(\beta_\ell(t_0,t_1|y_\ell))$ is not square summable and
hence time evolution cannot be unitarily implemented for any choice
of $SO(3)$-invariant complex structure.

\subsection{Conformal field redefinitions}

We will show now that we can avoid this negative conclusion much in
the same way as in the three-torus $\mathbb{T}^3$ case, i.e. by
introducing a redefinition of the fields in terms of which the model
is formulated \cite{Corichi:2006xi}. In our approach this
redefinition is suggested by the functional form of the conformal
factor $\sin t$ appearing in the auxiliary metric
$\mathring{g}_{ab}$ (\ref{g0}). In the following we will reintroduce
the index $i$ that labels the gravitational scalar ($i=0$) and the
matter scalars ($i=1,\dots,N$) and consider the
new fields
$$\xi_i:=\sqrt{\sin t}\phi_i\,.$$
The field equations are now
\begin{equation}
-\ddot{\xi}_i+\Delta_{\mathbb{S}^2}\xi_i=\frac{1}{4}(1+\csc^2t)\xi_i\,,\quad
\mathcal{L}_\sigma\xi_i=0\,. \label{(B)}
\end{equation}
They can be interpreted as the equation for a scalar, axially symmetric field with a time-dependent mass term $\frac{1}{4}(1+\csc^2t)$,  evolving in $(0,\pi)\times \mathbb{S}^2$ with the \textit{regular} --i.e. extensible to $\mathbb{R}\times \mathbb{S}^2$-- background metric
$$
\mathring{\eta}_{ab}=-(\mathrm{d}t)_a(\mathrm{d}t)_b+\gamma_{ab}\,.
$$
Notice that the mass term is singular at $t=0$ and $t=\pi$ but has
the \textit{correct sign} for all $t\in(0,\pi)$. This field
redefinition can be incorporated in the model at the Lagrangian
level by substituting $\phi_i=\xi_i/\sqrt{\sin t}$ in the action
(\ref{accion}) to get the corresponding variational problem in terms
of the new fields
\begin{eqnarray}
s(\xi_i)&=&-\frac{1}{2}\sum_{i=0}^N\int_{[t_0,t_1]\times \mathbb{S}^2} |\mathring{\eta}|^{1/2} \mathring{\eta}^{ab} \Big((\mathrm{d}\xi_i)_a(\mathrm{d}\xi_i)_b-(\mathrm{d}\log \sin t)_a(\mathrm{d}\xi_i)_b\xi_i\label{action_xi}\\
& & \hspace{5.9cm}+\frac{1}{4}(\mathrm{d}\log\sin t)_a(\mathrm{d}\log\sin t)_b \xi_i^2\Big)\,.\nonumber
\end{eqnarray}
We will follow now the method used in the preceding sections for the
original $\phi$ fields. Some details will be omitted owing to the
similarity with the previous derivations. Let us consider then the
space $\mathcal{S}_\xi$ of smooth and symmetric real solutions to
equation (\ref{(B)}) and expand $\xi\in\mathcal{S}_\xi$ as
\begin{equation}
\xi(t,s)=\sum_{\ell=0}^\infty\Big(b_\ell
z_\ell(t)Y_{\ell0}(s)+\overline{b_\ell
z_\ell(t)Y_{\ell0}(s)}\Big)\,,\label{fourier_xi}
\end{equation}
where $z_\ell(t)$ are complex functions satisfying the equations
\begin{equation}
\ddot{z}_\ell+\left(\frac{1}{4}(1+\csc^2t)+\ell(\ell+1)\right) z_\ell=0\,.
\label{(C)}
\end{equation}
The functions $z_\ell$ can be easily written in terms the functions $y_\ell$ appearing in (\ref{diffeqS3}) and satisfying (\ref{normaliz})
$$z_\ell(t)=\sqrt{\sin t} \,y_\ell(t)\,.$$
We immediately find that the Wronskian is now normalized to be
\begin{eqnarray}
z_\ell\dot{\bar{z}}_\ell-\bar{z}_\ell\dot{z}_\ell=i\,.\label{norm_zs}
\end{eqnarray}
This allows us to write the symplectic structure in
$\mathcal{S}_\xi$, derived from (\ref{action_xi}),  as
\begin{eqnarray*}
\Omega_\xi(\xi_1,\xi_2)&=&\int_{\mathbb{S}^2}|\gamma|^{1/2}\imath^*_t
\Big(\xi_2\dot{\xi}_1- \xi_1\dot{\xi}_2\Big)=
i\sum_{\ell=0}^\infty(\bar{b}_{1\ell}b_{2\ell}-\bar{b}_{2\ell}b_{1\ell})\,,
\quad\forall\xi_1,\xi_2\in\mathcal{S}_\xi\,.
\end{eqnarray*}

\noindent \textbf{Classical evolution}

\bigskip

We can consider now the classical functional time evolution operator
$\mathcal{T}_{(t_0,t_1)}:\Gamma_\xi\rightarrow\Gamma_\xi$
in the covariant phase space $\Gamma_\xi=(\mathcal{S}_\xi,\Omega_\xi)$. As before, we will write it in the form
\begin{eqnarray}
(\mathcal{T}_{(t_0,t_1)}\xi)(t,s):=\sum_{\ell=0}^\infty
\Big(\mathfrak{b}_\ell(t_0,t_1)
z_\ell(t)Y_{\ell0}(s)+\overline{\mathfrak{b}_\ell(t_0,t_1)
z_\ell(t)Y_{\ell0}(s)}\Big)\,. \label{eq1}
\end{eqnarray}
In this case, the map $\mathcal{T}_{(t_0,t_1)}=\mathfrak{J}_{t_0}\circ \mathfrak{J}_{t_1}^{-1}$ is constructed from
\begin{eqnarray}
\mathfrak{J}^{-1}_{t_1}:\Gamma_\xi\rightarrow \Upsilon\,,\quad \xi\mapsto(Q,P)=\mathfrak{J}^{-1}_{t_1}(\xi)\,,
\end{eqnarray}
defined by\footnote{Notice that the space of Cauchy data for the $\xi$-field equations is also $\Upsilon$.}
\begin{eqnarray}
Q(s)&:=&\xi(t_1,s)=\sum_{\ell=0}^\infty \Big( b_\ell z_\ell(t_1)Y_{\ell 0}(s)+ \overline{b_\ell z_\ell(t_1)Y_{\ell0}(s)}\Big)\label{Q_xi}\,,\\
P(s)&:=&\dot{\xi}(t_1,s)-\frac{1}{2}\cot t_1 \xi(t_1,s)\label{P_xi}\\
&=&
\sum_{\ell=0}^\infty \Big(b_\ell [\dot{z}_\ell(t_1)-\frac{1}{2}\cot t_1 z_\ell(t_1)]Y_{\ell0}(s)+\overline{b_\ell [\dot{z}_\ell(t_1)-\frac{1}{2}\cot t_1 z_\ell(t_1)]Y_{\ell0}(s)}\Big)\,,\nonumber
\end{eqnarray}
and from
\begin{eqnarray}
\mathfrak{J}_{t_0}: \Upsilon\rightarrow \Gamma_\xi\,,\quad (Q,P)\mapsto  \xi=\mathfrak{J}_{t_0}(Q,P)
\end{eqnarray}
defined, in terms of the Fourier coefficients $b_\ell$ of $\xi$
(\ref{fourier_xi}), by
$$
b_\ell(t_0)=- i[\dot{\bar{z}}_\ell(t_0)-\frac{1}{2}\cot t_0
\bar{z}_\ell(t_0)]\int_{\mathbb{S}^2}|\gamma|^{1/2}Y_{\ell0} Q +i
\bar{z}_\ell(t_0)\int_{\mathbb{S}^2}|\gamma|^{1/2}Y_{\ell0}P\,.
$$
From these expressions we obtain
\begin{eqnarray}
\mathfrak{b}_\ell(t_0,t_1)&=&-i\Big[z_\ell(t_1)\Big(\dot{\bar{z}}_\ell(t_0)-\frac{1}{2}\cot t_0\bar{z}_\ell(t_0)\Big)- \bar{z}_\ell(t_0)\Big(\dot{z}_\ell(t_1)-\frac{1}{2}\cot t_1 z_\ell(t_1)\Big)\Big]\, b_\ell\phantom{espacio}\label{eq2}\\
&&-i\Big[\bar{z}_\ell(t_1)\Big(\dot{\bar{z}}_\ell(t_0)-\frac{1}{2}\cot t_0 \bar{z}_\ell(t_0)\Big)- \bar{z}_\ell(t_0)\Big(\dot{\bar{z}}_\ell(t_1)-\frac{1}{2}\cot t_1 \bar{z}_\ell(t_1)\Big)\Big]\,
\bar{b}_\ell\,.\nonumber
\end{eqnarray}

\noindent \textbf{Quantum evolution}

\bigskip

We will analyze the continuity of the symplectic transformation defined by (\ref{eq1}) and (\ref{eq2}) at the end of this section and consider first the unitarity condition for the quantum evolution in the corresponding Fock space quantization
\begin{eqnarray}
\sum_{\ell=0}^\infty|\beta^\xi_\ell(t_0,t_1|z_\ell)|^2=\sum_{\ell=0}^\infty\Big(
\mathrm{Re}^2(\beta^\xi_\ell(t_0,t_1|z_\ell))
+\mathrm{Im}^2(\beta^\xi_\ell(t_0,t_1|z_\ell))\Big)
<\infty\,, \label{(D)}
\end{eqnarray}
for all $t_0$,$t_1\in(0,\pi)$, where
\begin{eqnarray}
\beta^\xi_\ell(t_0,t_1|z_\ell):=z_\ell(t_1)
\Big(\dot{z}_\ell(t_0)-\frac{1}{2}\cot
t_0z_\ell(t_0)\Big)-z_\ell(t_0)
\Big(\dot{z}_\ell(t_1)-\frac{1}{2}\cot
t_1z_\ell(t_1)\Big)\,.\label{betaxi}
\end{eqnarray}
The general solution of equation (\ref{(C)}) with the normalization (\ref{norm_zs}) can be written, again, in terms of associated Legendre functions (\ref{u0v0}) in the form
\begin{eqnarray*}
 z_\ell(t)&=&\rho_\ell \sqrt{\sin t}\, u_{0\ell}(t)+(\nu_\ell+i\rho^{-1}_\ell)\sqrt{\sin t}\, v_{0\ell}(t)\\
 &=& \rho_\ell \tilde{u}_{0\ell}(t)+(\nu_\ell+i\rho^{-1}_\ell)\tilde{v}_{0\ell}(t)\,,
\end{eqnarray*}
where, as above, $\rho_\ell>0$ and $\nu_\ell\in \mathbb{R}$
parameterize different choices of one-particle Hilbert spaces, and
we have defined $\tilde{u}_{0\ell}:=\sqrt{\sin t}u_{0\ell}$ and
$\tilde{v}_{0\ell}:=\sqrt{\sin t}v_{0\ell}$.
 We have to discuss now the convergence condition expressed in (\ref{(D)}). Let us consider first
\begin{eqnarray*}
\mathrm{Im}(\beta^\xi_\ell(t_0,t_1|z_\ell))&=&\tilde{A}_\ell(t_{0},t_{1})
+2\nu_{\ell}\rho_{\ell}^{-1}\tilde{B}_{\ell}(t_{0},t_{1})
\end{eqnarray*}
where
\begin{eqnarray*}
\tilde{A}_\ell(t_{0},t_{1})&:=& \tilde{u}_{0\ell}(t_{1})\dot{\tilde{v}}_{0\ell}(t_{0})-\tilde{u}_{0\ell}(t_{0})\dot{\tilde{v}}_{0\ell}(t_{1})
+\dot{\tilde{u}}_{0\ell}(t_{0})\tilde{v}_{0\ell}(t_{1})-\dot{\tilde{u}}_{0\ell}(t_{1})\tilde{v}_{0\ell}(t_{0})\\
&+& \frac{1}{2}(\cot t_{1}-\cot t_{0})\left(\tilde{u}_{0\ell}(t_{1})\tilde{v}_{0\ell}(t_{0})+\tilde{u}_{0\ell}(t_{0})\tilde{v}_{0\ell}(t_{1})\right)\,,\\
\tilde{B}_{\ell}(t_{0},t_{1})&:=&\tilde{v}_{0\ell}(t_{1})\dot{\tilde{v}}_{0\ell}(t_{0})
-\tilde{v}_{0\ell}(t_{0})\dot{\tilde{v}}_{0\ell}(t_{1})
+\frac{1}{2}(\cot t_{1}-\cot t_{0})\tilde{v}_{0\ell}(t_{0})\tilde{v}_{0\ell}(t_{1})\,.
\end{eqnarray*}
The asymptotic behavior of $\tilde{A}_\ell$ and $\tilde{B}_\ell$ as
$\ell\rightarrow\infty$ can be obtained from (\ref{asintoticoQP})
and (\ref{u0v0})
\begin{eqnarray}
\tilde{A}_\ell(t_{0},t_{1})& \sim & \frac{1}{4\ell}(\cot t_{1}-\cot
t_{0})\cos[(\ell+1/2)(t_{0}+t_{1})]\,, \label{asint_A}\\
\tilde{B}_\ell(t_{0},t_{1})& \sim &  \frac{\pi}{4}\sin[(\ell+1/2)(t_{1}-t_{0})]\,.\label{asint_B}
\end{eqnarray}
We then conclude that $\mathrm{Im}(\beta_\ell^\xi(t_0,t_1|z_\ell))$ is square summable if $\nu_\ell\rho^{-1}_\ell\sim\ell^{-a}$ (with $a\geq 1$) when $\ell\rightarrow\infty$. For the real part we have
\begin{eqnarray*}
\mathrm{Re}(\beta^\xi_\ell(t_0,t_1|z_\ell))&=&\rho_{\ell}\nu_{\ell}
\tilde{A}_{\ell}(t_{0},t_{1})+(\nu_{\ell}^{2}-\rho_{\ell}^{-2})\tilde{B}_{\ell}(t_{0},t_{1})
+\rho_{\ell}^{2}\tilde{C}_{\ell}(t_{0},t_{1})\,,
\end{eqnarray*}
where
\begin{eqnarray}
\tilde{C}_{\ell}(t_{0},t_{1})&:=&\tilde{u}_{0\ell}(t_{1})\dot{\tilde{u}}_{0\ell}(t_{0})
-\tilde{u}_{0\ell}(t_{0})\dot{\tilde{u}}_{0\ell}(t_{1})
+\frac{1}{2}(\cot t_{1}-\cot t_{0})\tilde{u}_{0\ell}(t_{0})\tilde{u}_{0\ell}(t_{1})\nonumber\\
& \sim & \frac{1}{\pi}\sin[(\ell+1/2)(t_{1}-t_{0})],\,\,\,
\mathrm{when}\,\,\, \ell\rightarrow\infty. \label{asint_C}
\end{eqnarray}
The asymptotic behavior as $\ell\rightarrow\infty$ of $\mathrm{Re}(\beta^\xi_\ell(t_0,t_1|z_\ell))$ can be obtained now from (\ref{asint_A}), (\ref{asint_B}), and (\ref{asint_C}). If we choose now $\rho_\ell$ in such a way that
\begin{equation}
\rho_\ell\sim\sqrt{\frac{\pi}{2}} \quad  \textrm{and} \quad \nu_\ell\sim\ell^{-a}\quad  (a\geq1)\quad \textrm{as}\quad  \ell\rightarrow\infty  \label{condiciones}
\end{equation}
we also guarantee that $\mathrm{Re}(\beta_\ell^\xi(t_0,t_1|z_\ell))$
is square summable, and hence $\beta_\ell^\xi(t_0,t_1|z_\ell)$.

\bigskip

We end this section by showing that the linear symplectic map
$\mathcal{T}_{(t_0,t_1)}$ is continuous in the norm $||\cdot||=\sqrt{\langle\cdot|\cdot\rangle|_{\mathcal{P}}}$ associated to the inner product (\ref{inprod}) for the complex structures
characterized by the pairs $(\rho_\ell,\nu_\ell)$ verifying
(\ref{condiciones}). That is, there exists some
$K(t_0,t_1)>0$ such that
$$\|\kappa(\mathcal{T}_{(t_0,t_1)}\xi)\|\leq K(t_0,t_1)\|\kappa(\xi)\|$$
for all $\xi\in \mathcal{S}_\xi$, where
$\kappa:\mathcal{S}_{\xi_{\mathbb{C}}}\rightarrow\mathcal{P}_{\xi}$
is the $\mathbb{C}$-linear projector defined by the splitting
$\mathcal{S}_{\xi_\mathbb{C}}=\mathcal{P}_\xi\oplus\bar{\mathcal{P}}_\xi$.
By using (\ref{eq1}) and (\ref{eq2}) it is straightforward to show
that
\begin{eqnarray}
\|\kappa(\mathcal{T}_{(t_0,t_1)}\xi)\|^2=\sum_{\ell=0}^\infty
|\mathfrak{b}_\ell(t_0,t_1)|^2\leq \sum_{\ell=0}^\infty
\Big(|\alpha_\ell(t_0,t_1|z_\ell)|^2+|\beta^\xi_\ell(t_0,t_1|z_\ell)|^2\Big)
|b_\ell|^2\label{cont}
\end{eqnarray}
where
\begin{eqnarray*}
\alpha^\xi_\ell(t_0,t_1|z_\ell)&:=& z_\ell(t_1)
\Big(\dot{\bar{z}}_\ell(t_0)-\frac{1}{2}\cot t_0 \bar{z}_\ell(t_0)
\Big)-\bar{z}_\ell(t_0) \Big(\dot{z}_\ell(t_1)-\frac{1}{2}\cot t_1
z_\ell(t_1)\Big)\,,
\end{eqnarray*}
and $\beta^\xi_\ell(t_0,t_1|z_\ell)$ is given by (\ref{betaxi}). We
have shown above that the sequence
$\{|\beta^\xi_\ell(t_0,t_1|z_\ell)|^2\}_{\ell=0}^{\infty}$ is
bounded (actually square summable) so if we can see now that
$\{|\alpha^\xi_\ell(t_0,t_1|z_\ell)|^2\}_{\ell=0}^{\infty}$ is also
a bounded sequence the continuity of $\mathcal{T}_{(t_0,t_1)}$
follows directly from equation (\ref{cont}). By expanding $z_\ell=
\rho_\ell\tilde{u}_{0\ell}+(\nu_\ell+i\rho^{-1}_\ell)\tilde{v}_{0\ell}$
--and making use of (\ref{asympt}), (\ref{asint_A}),
(\ref{asint_B}), (\ref{asint_C}), and (\ref{condiciones})-- it is
possible to show that
\begin{eqnarray*}
\mathrm{Re}(\alpha^\xi_\ell(t_0,t_1|z_\ell))&=& \rho^2_\ell \tilde{C}_\ell(t_0,t_1) + \rho_\ell\nu_\ell \tilde{A}_\ell(t_0,t_1) + (\nu^2_\ell+\rho^{-2}_\ell) \tilde{B}_\ell(t_0,t_1)\\
&\sim& \sin[(\ell+1/2)(t_1-t_0)] \quad \textrm{when} \quad
\ell\rightarrow \infty\,,
\\
\mathrm{Im}(\alpha^\xi_\ell(t_0,t_1|z_\ell))&=&\tilde{v}_{0\ell}(t_1)\dot{\tilde{u}}_{0\ell}(t_0)
-\tilde{u}_{0\ell}(t_0)\dot{\tilde{v}}_{0\ell}(t_1)
+\tilde{v}_{0\ell}(t_0)\dot{\tilde{u}}_{0\ell}(t_1)
-\tilde{u}_{0\ell}(t_1)\dot{\tilde{v}}_{0\ell}(t_0)
\\
&+&\frac{1}{2} (\cot t_0-\cot t_1)\Big(
\tilde{u}_{0\ell}(t_1)\tilde{v}_{0\ell}(t_0)
-\tilde{u}_{0\ell}(t_0)\tilde{v}_{0\ell}(t_1)
\Big)\\
&\sim & \cos[(\ell+1/2)(t_1-t_0)] \quad \textrm{when} \quad \ell\rightarrow \infty\,.
\end{eqnarray*}
From these equations it is clear that there exists a $K^2(t_0,t_1)>0$ such that
$$
|\alpha^\xi_\ell(t_0,t_1|z_\ell)|^2+|\beta^\xi_\ell(t_0,t_1|z_\ell)|^2\leq K^2(t_0,t_1)\,, \quad \forall \ell\in \mathbb{N}\cup\{0\}\,.
$$
Then, using (\ref{cont}), we get that
$\|\kappa(\mathcal{T}_{(t_0,t_1)}\xi)\|^2\leq
K^2(t_0,t_1)\|\kappa(\xi)\|^2$, and hence $\mathcal{T}_{(t_0,t_1)}$
is continuous. In conclusion, by imposing suitable conditions
(\ref{condiciones}) on the parameters $\rho_\ell$ and $\nu_\ell$, it
is possible to find $SO(3)$-complex structures (and, hence, subspaces
$\mathcal{P}$) such that the quantum dynamics can be unitarily
implemented in $\mathcal{F}_s(\mathcal{H}_{\mathcal{P}})$.

\subsection{Uniqueness of the Fock quantization}{\label{uniqueness}}

We will show in this section that any two Fock quantizations of the
field $\xi$ corresponding to $SO(3)$-invariant complex structures,
for which the dynamics can be unitarily implemented, are equivalent.
To this end, let us recall some properties of the $SO(3)$-invariant
complex structures considered in section \ref{Invariant complex
forms}. Given any invariant complex structure $J$, it is possible to
characterize its action on the fixed basis $\phi_{0\ell}$ that
defines the complex structure $J_0$. This action is given by
equation (\ref{JJ0}). As we can see there exists a linear symplectic
transformation $T_{J}$ connecting them, so that $J=T_{J}\circ
J_{0}\circ T_{J}^{-1}$. Explicitly
\begin{equation}
T_{J}=\bigoplus_{\ell=0}^{\infty}\left(\begin{array}{cc}
(\tau_{1}^{\ell})_{J}I_{11}^{\ell}&(\tau_{2}^{\ell})_{J}I_{12}^{\ell}\\
(\bar{\tau}_{2}^{\ell})_{J}I_{21}^{\ell}&(\bar{\tau}_{1}^{\ell})_{J}I_{22}^{\ell}\end{array}\right),
\end{equation}
with
\begin{eqnarray*}
(\tau_{1}^{\ell})_{J}&:=&\sqrt{(1+|\jmath^\ell_{11}|)/2}\quad \textrm{(up to
multiplicative phase)}\,,\\
(\tau_{2}^{\ell})_{J}&:=&\frac{i\jmath^\ell_{12}}{2(\tau_{1}^{\ell})_J}\,.
\end{eqnarray*}
Notice that $J_0$, defined by the set of functions
$\{z_{0\ell}(t)=\tilde{u}_{0\ell}(t)+i\tilde{v}_{0\ell}\}_{\ell=0}^{\infty}$, corresponding to $\rho_{\ell}=1$ and $\nu_{\ell}=0$, does not lead to a unitary
implementation of dynamics. In this context, it is fixed just to
compare different complex structures. Let us consider then any two
$SO(3)$-invariant complex structures, $J$ and $J'$, for which the
dynamics is unitary. They will define unitarily equivalent quantum
theories if and only if the linear symplectic transformation
$T_{J,J'}:=T_{J}\circ T_{J'}^{-1}$ connecting them through
$J=T_{J,J'}\circ J'\circ T_{J,J'}^{-1}$ is unitarily implementable.
This is the case if the sequence
$$
\{(\tau_{2}^{\ell})_{J}(\tau_{1}^{\ell})_{J'}
-(\tau_{1}^{\ell})_{J}(\tau_{2}^{\ell})_{J'}\}_{\ell=0}^{\infty}
$$
is square summable. Taking into account the relations (\ref{j1}) and
(\ref{j2}), as well as the asymptotic behaviors (\ref{condiciones}), the previous
condition is indeed verified, so the quantum theories defined by $J$
and $J'$ are unitarily equivalent.

\subsection{Normalizability of the action of the Hamiltonian on the vacuum state}

We discuss here an interesting feature of the quantum dynamics for
these systems: The fact that, even though the evolution is unitarily
implemented, the time-dependent quantum Hamiltonian, constructed
from the classical one by following the standard rules of
quantization, has the striking property  that Fock space vectors
corresponding to a finite number of particle-like excitations do not
belong to its domain. This also happens in the $\mathbb{T}^3$ case \cite{Corichi:2006xi}.

The classical Hamiltonian on the canonical phase space $\Upsilon$ in
the $\xi$-description of the system is derived from the action
(\ref{action_xi}). It is given by
\begin{equation}
H(Q,P;t)=\frac{1}{2}\int_{\mathbb{S}^2}|\gamma|^{1/2}\left(P^2+\cot
t \,PQ-Q\Delta_{\mathbb{S}^2}Q\right). \label{Hamiltoniano}
\end{equation}
Notice that the time-dependent (non-autonomous) Hamiltonian (\ref{Hamiltoniano}) is
an \textit{indefinite} quadratic form with a cross term involving
$Q$ and $P$. Let us discuss now the quantum Hamiltonian. To this end
we first write the formal quantum version of (\ref{Q_xi}) and
(\ref{P_xi}) that should be understood as operador-valuated
distributions on $\mathbb{S}^{2}$ for each value of $t$
\begin{eqnarray}
\hat{Q}(t,s)&:=& \sum_{\ell=0}^\infty \Big( z_\ell(t)Y_{\ell0}(s)\,
\hat{b}_\ell+
 \overline{z_\ell(t)Y_{\ell0}(s)}\,\hat{b}^\dag_\ell\Big)\,,\nonumber\\
\hat{P}(t,s)&:=& \sum_{\ell=0}^\infty \Big(
[\dot{z}_\ell(t)-\frac{1}{2}\cot t z_\ell(t)]Y_{\ell0}(s)\,
\hat{b}_\ell+ \overline{[\dot{z}_\ell(t)-\frac{1}{2}\cot t
z_\ell(t)]Y_{\ell0}(s)}\,\hat{b}^\dag_\ell\Big)\,,\nonumber
\end{eqnarray}
where $\hat{b}_\ell$ and $\hat{b}_\ell^\dagger$ are the annihilation
and creation operators associated to the modes $\xi_\ell=z_\ell
Y_{\ell0}$, respectively. Substituting these expressions in
(\ref{Hamiltoniano}), and after normal ordering, we find
\begin{eqnarray}
\hat{H}(t)&=&\frac{1}{2}\sum_{\ell=0}^\infty \Big( K_\ell(t)\,\hat{b}_\ell^2 +\bar{K}_\ell(t)\,\hat{b}_\ell^{\dag2}
+2G_\ell(t)\,\hat{b}_\ell^\dag \hat{b}_\ell\Big)\,,
\label{qham}
\end{eqnarray}
where
\begin{eqnarray}
K_\ell(t)&:=&\big(\dot{z}_\ell(t)-\frac{1}{2}\cot t z_\ell(t)\big)^2+\ell(\ell+1)z_\ell^2(t)+\cot t \big(\dot{z}_\ell(t)-\frac{1}{2}\cot t z_\ell(t)\big) z_\ell(t)\,,\label{K_ell}\\
G_\ell(t)&:=&|\dot{z}_\ell(t)-\frac{1}{2}\cot t
z_\ell(t)|^2+\ell(\ell+1)|z_\ell(t)|^2\nonumber\\
&&+\frac{1}{2}\cot t\Big(\big(\dot{z}_\ell(t)-\frac{1}{2}\cot t
z_\ell(t)\big)
\bar{z}_\ell(t)+\big(\dot{\bar{z}}_\ell(t)-\frac{1}{2}\cot t
\bar{z}_\ell(t)\big)z_\ell(t)\Big).\nonumber
\end{eqnarray}
The action of the quantum Hamiltonian on the vacuum
$|0\rangle$ is now
$$
\hat{H}(t)|0\rangle=\frac{1}{\sqrt{2}}\sum_{\ell=0}^\infty
\bar{K}_\ell(t)|2_\ell\rangle\,,
$$
where
$\sqrt{2}|2_\ell\rangle=\hat{b}_\ell^{\dag 2}\,|0\rangle$.
The state $\hat{H}(t)|0\rangle$ will be normalizable if and only if
\begin{equation}
\sum_{\ell=0}^\infty|K_\ell(t)|^2<\infty\,. \label{(E)}
\end{equation}
Taking into account the asymptotic behaviors of the Legendre
functions (\ref{asintoticoQP}) when $\ell\rightarrow\infty$, and
imposing the conditions $\rho_\ell\sim \sqrt{\pi/2}$ and
$\nu_\ell\sim \ell^{-a}$ discussed above to guarantee the unitary
implementation of the time evolution, we get
\begin{eqnarray*}
& & z_\ell(t)=\rho_\ell \sqrt{\sin t}\, u_{0\ell}(t)+(\nu_\ell+i\rho^{-1}_\ell)\sqrt{\sin t}\, v_{0\ell}(t)\sim\frac{1}{\sqrt{2\ell}}\exp\Big(-i[(\ell+1/2)t-\pi/4]\Big)\,,\\
& & \dot{z}_\ell(t)-\frac{1}{2}\cot t z_\ell(t)\sim-i\sqrt{\frac{\ell}{2}}\exp\Big(-i[(\ell+1/2)t-\pi/4]\Big)\,.
\end{eqnarray*}
It is straightforward now to compute the asymptotic behavior of $K_\ell(t)$ defined in (\ref{K_ell}) and  also check that condition (\ref{(E)}) is not satisfied. We then conclude that the Fock space vacuum does not belong to the domain of the Hamiltonian for any time $t\in(0,\pi)$ and, hence, the action of the Hamiltonian on $n$-particle states is not defined either.

It is important to point out that it is possible to consider the
definition of the quantum Hamiltonian in a more mathematical
framework. It is well known that the unitary evolution operator
$\hat{U}(t_0, t_1)$ can be derived from the evolution of creation
operators in the Heisenberg picture and the evolution of the vacuum
state. Furthermore, the vacuum evolution can be written in closed
form as in \cite{Torre:1998,Pilch} and is given by a completely analogous formula. As expected in a non-autonomous system, the vacuum state (and, hence, $n$-particle states) is not stable under time evolution. After computing the explicit
form of the evolution operator, it is possible to  study the
differentiability of  $\hat{U}(t_0, t_1)$ in a rigorous mathematical
sense and then, whenever $\hat{U}$ is differentiable, we can define
the quantum Hamiltonian of the system. This is beyond the scope of
the present paper.

We end this section by noting that the covariant phase space $\Gamma_\xi$ defined by (\ref{action_xi}) can be equivalently derived from the simpler action
\begin{eqnarray}
s_0(\xi)&=&-\frac{1}{2}\int_{[t_0,t_1]\times \mathbb{S}^2} |\mathring{\eta}|^{1/2} \mathring{\eta}^{ab} \Big((\mathrm{d}\xi)_a(\mathrm{d}\xi)_b+\frac{1}{4}(1+\csc^2t) \,\xi^2\Big)\,.\label{action_xi0}
\end{eqnarray}
This variational principle gives a time-dependent, \textit{positive definite}, diagonal Hamiltonian of the form
\begin{eqnarray*}
H_0(Q,P;t)=\frac{1}{2}\int_{\mathbb{S}^2}|\gamma|^{1/2}\left(P^2+Q\Big[\frac{1}{4}(1+\csc^2t)-\Delta_{\mathbb{S}^2}\Big]Q\right).
\label{Hamiltoniano0}
\end{eqnarray*}
There are no subtleties associated to the domain of  the quantum counterpart of $H_0$ in the
sense that now the Fock space vacuum belongs to the domain of the
Hamiltonian. The action principle (\ref{action_xi0}) is related to the Einstein-Hilbert action for the Gowdy models (\ref{accion}) through a field redefinition. In fact both actions can be connected by a time-dependent canonical transformation though nothing guarantees that this can be unitarily implemented, in which case the quantizations would be different.

\section{Conclusions and comments}{\label{conclusions}}

As we have shown in the paper there is a very natural framework to
discuss issues related to the unitary implementability of dynamics
in the compact Gowdy models. The key idea is to use a covariant
phase space approach where the solutions to the field equations play
the main role. The best way to  describe these solution spaces
\cite{BarberoG.:2007} is by rewriting the field equations in terms
of certain auxiliary background metrics that are conformally
equivalent to some specially simple and natural ones.  For the
$\mathbb{T}^3$ case, this metric is the flat metric on
$(0,\infty)\times\mathbb{T}^2$, and for the
$\mathbb{S}^1\times\mathbb{S}^2$ and $\mathbb{S}^3$ examples the metric
is the Einstein metric on $(0,\pi)\times\mathbb{S}^2$. It is
important to highlight the fact that this is possible as a
consequence of the symmetry left in the model after its reduction to
$(1+2)$-dimensions. This symmetry is generated by the Killing field
remaining after the Geroch reduction from $(1+3)$ to $(1+2)$
dimensions. An advantage of this approach is the fact that the time
singularities of the metric are completely described by the
time-dependent conformal factors. The metric becomes singular
whenever they cancel. This ultimately explains why a simple field
redefinition involving \textit{precisely} these conformal factors
suffices to cure the problems associated with the quantum unitary
evolution. In fact a conformal transformation defined with the help
of these conformal factors shifts the singularity of the metric to
one appearing in a time-dependent potential term that becomes
singular when the full metric does.

A first result of the paper is a proof of the fact that the
impossibility to get unitary dynamics in terms of the original
variables that naturally appear in the description of the model is
insensitive to the choice of the complex structure used in the
quantization. This result generalizes the conclusion reached in
\cite{Corichi:2006xi} for the $\mathbb{T}^3$ case to the topologies
considered here ($\mathbb{S}^1\times{\mathbb{S}}^2$ and
$\mathbb{S}^3$). The starting point of the approach that we develop
in the paper is to consider the possibility of achieving unitary
quantum evolution by making an appropriate choice of complex
structure; only when this fails are we forced to introduce new
variables to describe the system\footnote{For the $\mathbb{T}^3$
case this is obtained as a corollary of the uniqueness result
described in \cite{Mena:2007}.}. It is interesting to point out in
this respect that the type of unitarity problem discussed here
cannot always be fixed by time-dependent redefinitions of the type
used in the paper; in fact it is possible to give examples (a
massless scalar field evolving in a de Sitter background) where this
is not the case \cite{EdDan}. The ultimate reason why the method
used here does not work in these other models is the fact that the
time-dependent potential written in terms of the new fields is not
as well behaved as the ones that show up in the treatment of the
Gowdy models.

A second point that we want to comment on is the uniqueness issue.
In the case of the $\mathbb{T}^3$ Gowdy models the presence of a
constraint, and the corresponding symmetry generated by it, gives
the possibility of introducing a physically sensible criterion to
select the complex structure: invariance under this symmetry
\cite{Corichi:2006zv}. This is not the case for the other compact
topologies that we consider here for which, as we showed in
\cite{BarberoG.:2007}, there are no extra constraints after
deparameterization. It is important to realize in this respect that 
we have used the $SO(3)$ symmetry associated to the background metric to select a preferred class of complex structures.

Notice that at this point we still have many different $SO(3)$-invariant Fock quantizations $\mathcal{F}_s(\mathcal{H}_{\mathcal{P}})$ labeled by $\mathcal{P}$ that, in principle,
are not guaranteed to be equivalent. In such a situation we would need an additional
criterion to pick one. Once we require that the quantum dynamics is unitary we find that all of them are unitarily equivalent.

A final comment is to note that the same scheme followed here works
in the $\mathbb{T}^3$ case (with or without massless scalar matter).
For the vacuum case one directly recovers several interesting
results discussed in the literature for this system.

\begin{acknowledgments}
Daniel G\'omez Vergel acknowledges the support of
the Spanish Research Council (CSIC) through a I3P research
assistantship. This work is also supported by the Spanish MEC under
the research grant FIS2005-05736-C03-02.

\end{acknowledgments}

\end{document}